\documentclass[prb,aps,twocolumn,showpacs,nobibnotes,epsf]{revtex4}

\usepackage{graphicx}
\usepackage{dcolumn}
\usepackage{bm}
\usepackage{SIunits}

\begin{document}
\title{Study of Electron Spin Resonance on single crystals $EuFe_{2-x}Co_{x}As_2$}
\author{J. J. Ying, T. Wu, Q. J. Zheng, Y. He, G. Wu, Q. J. Li, Y. J. Yan, Y. L. Xie, R. H. Liu, X. F. Wang}
\author{X. H. Chen}
\altaffiliation{Corresponding author} \email{chenxh@ustc.edu.cn}
\affiliation{Hefei National Laboratory for Physical Science at
Microscale and Department of Physics, University of Science and
Technology of China, Hefei, Anhui 230026, People's Republic of
China\\}

\begin{abstract}
The temperature dependence of electron spin resonance (ESR) was
studied in $EuFe_{2-x}Co_{x}As_2 $ (x = 0.0 ,0.067 ,0.1 ,0.2 ,0.25
,0.285 ,0.35 ,0.4 and 0.5). The ESR spectrum of all the samples
indicates that the g factor and peak-to-peak linewidth strongly
depend on the temperature. Moreover, the peak-to-peak linewidth
shows the Korringa behavior, indicating an exchange coupling between
the conduction electrons and the $Eu^{2+}$ions. The linewidth, g
factor and the integrate ESR intensity show anomalies at the
temperature of the spin-density-wave (SDW). The linewidth below the
SDW transition does not rely on the temperature. This gives the
evidence of the gap opening at the $T_{SDW}$. The slope of the
linewidth is closely associated to $T_{SDW}$and $T_C$. This exotic
behavior may be related to the nesting of the Fermi surface.
\end{abstract}


\vskip 300 pt

\maketitle The discovery of Fe-based high Tc superconductor provides
a new family of materials to explore the mechanism of high-Tc
superconductivity besides high-Tc cuprates
superconductor\cite{yoichi, chenxh, chen, ren, rotter}. Ternary iron
arsenides $AFe_2As_2$ (A=Sr,Ca,Ba,Eu) is one of parent compounds
with $ThCr_2Si_2$-type structure. In analogy with LnFeAsO
(Ln=La-Gd), $AFe_2As_2$ undergoes a structural phase transition and
a spin-density-wave (SDW) transition, accompanied by the anomalies
in electrical resistivity, magnetic susceptibility and specific
heat. With doping electron or hole, the ground state of FeAs
compounds evolves from SDW state to superconducting (SC) state.

$EuFe_2As_2$ is a unique member in the ternary iron arsenide family
due to the large local moment of $Eu^{2+}$, which orders
antiferromagnetically below 20K\cite{zhi ren,wu}. Except for this
AFM transition, the physical properties of $EuFe_2As_2$ are found to
be quite similar with those of its isostructural compounds such as
$BaFe_2As_2$ and $SrFe_2As_2$. Wu et al. have gave the evidence for
a coupling between magnetism of $Eu^{2+}$ ions and SDW ordering in
FeAs layer\cite{wu}. Study on the interaction between the conduction
electrons of FeAs layer and localized spins of $Eu^{2+}$ layer gives
significant information about the properties of FeAs layer. The ESR
spectra in metals are very effective to study the interaction with
the conduction electrons and local moments, which has been applied
in high-Tc cuprates\cite{shaltiel,kataev}.

In this paper, we systematically studied the ESR spectra of the
single crystals $EuFe_{2-x}Co_{x}As_2 $(x=0, 0.067, 0.1, 0.2, 0.25,
0.285, 0.35, 0.4 and 0.5). The g factor, peak-to-peak linewidth and
the integrated ESR intensity all show anomalies at the $T_{SDW}$.
The linewidth of all the samples shows the Korringa behavior above
the $T_{SDW}$, indicating an exchange coupling between the
conduction electrons and spin of the $Eu^{2+}$ ions. The fact that
linewidth becomes temperature independence below $T_{SDW}$ gives the
evidence of the gap opening at $T_{SDW}$. The slope of the linewidth
shows interesting behavior with Co doping, and is closely related
with the characteristic temperature: the $T_{SDW}$ and the $T_c$.

\begin{figure}[t]
\centering
\includegraphics[width=0.5 \textwidth]{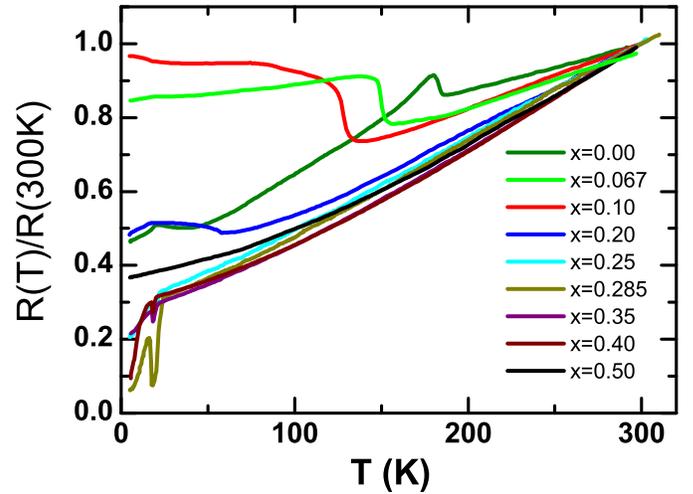}
\caption{(Color online) Temperature dependence of in-plane
resistivity for the single crystals $EuFe_{2-x}Co_{x}As_2$ with x=0,
0.067, 0.1, 0.2, 0.25, 0.285, 0.35, 0.4 and 0.5. } \label{fig1}
\end{figure}

High quality single crystals with nominal composition
$EuFe_{2-x}Co_{x}As_2 $(x=0 ,0.067 ,0.1 ,0.2 ,0.25 ,0.285 ,0.35 ,0.4
and 0.5) were grown by self-flux method as described for growth of
$BaFe_2As_2$ single crystals with FeAs as flux\cite{wang}. Many
shinning plate-like $EuFe_{2-x}Co_{x}As_2$ crystals were obtained.
The ESR measurements of the single crystals were performed using a
Bruker ER-200D-SRC spectrometer, operating at X-band frequencies
(9.07 GHz) and between 110K and 300K. The resistance was measured by
an AC resistance bridge (LR-700, Linear Research). It should be
addressed that all results discussed as follows are well
reproducible. We put the ab plane of single crystals perpendicular
to the magnetic field direction.

Fig.1 shows the temperature dependence of in-plane resistivity for
single crystals $EuFe_{2-x}Co_{x}As_2$ with x=0 ,0.067 ,0.1 ,0.2
,0.25 ,0.285 ,0.35 ,0.4 and 0.5. For parent compound $EuFe_2As_2$,
the resistivity shows a steep increase around 188 K, and reaches its
maximum value around 180 K. It is associated with SDW and structure
transition. The kink at about 20 K is related to antiferromagnetic
ordering of $Eu^{2+}$ moments. With Co doping, the SDW transition is
gradually suppressed. For the crystal with x=0.25, the SDW
transition was completely suppressed and shows superconducting
transition at $T_c \sim 22$ K,  but the resistance could not reach
zero because of the antiferromagnetic ordering of the large local
moment of $Eu^{2+}$ ions. This issue is studied in separated
work.\cite{zheng} For the optimally doped crystal with x=0.285, the
Tc reaches 24.5 K. For x=0.35 and x=0.4 samples, $T_c$ decreases to
21 K and 20 K, respectively, while neither the SDW transition nor
the superconductivity is observed for the heavily doped crystal with
x=0.5.

Fig.2(a) and (b) show the temperature dependence of ESR spectra for
the parent compound $EuFe_2As_2$ and one of the selected Co-doped
samples $EuFe_{1.933}Co_{0.067}As_2$ in the temperature ranging from
110 K to 300 K, respectively. A well defined paramagnetic signal was
observed in the whole temperature range. We ascribe the signal to
the Eu ions because only weak signal can be detected in the
$BaFe_2As_2$ single crystal. The intensity of the spectra is
obviously suppressed with increasing temperature. The spectrum
abruptly becomes narrow and the intensity is enhanced around
$T_{SDW}$ with increasing temperature. With further increasing
temperature, the spectra broadens and the intensity decreases. The
detailed analysis of the spectrum is discussed later on. In other
Co-doped compounds, the similar signal has also been observed.

\begin{figure}[t]
\centering
\includegraphics[width=0.5\textwidth]{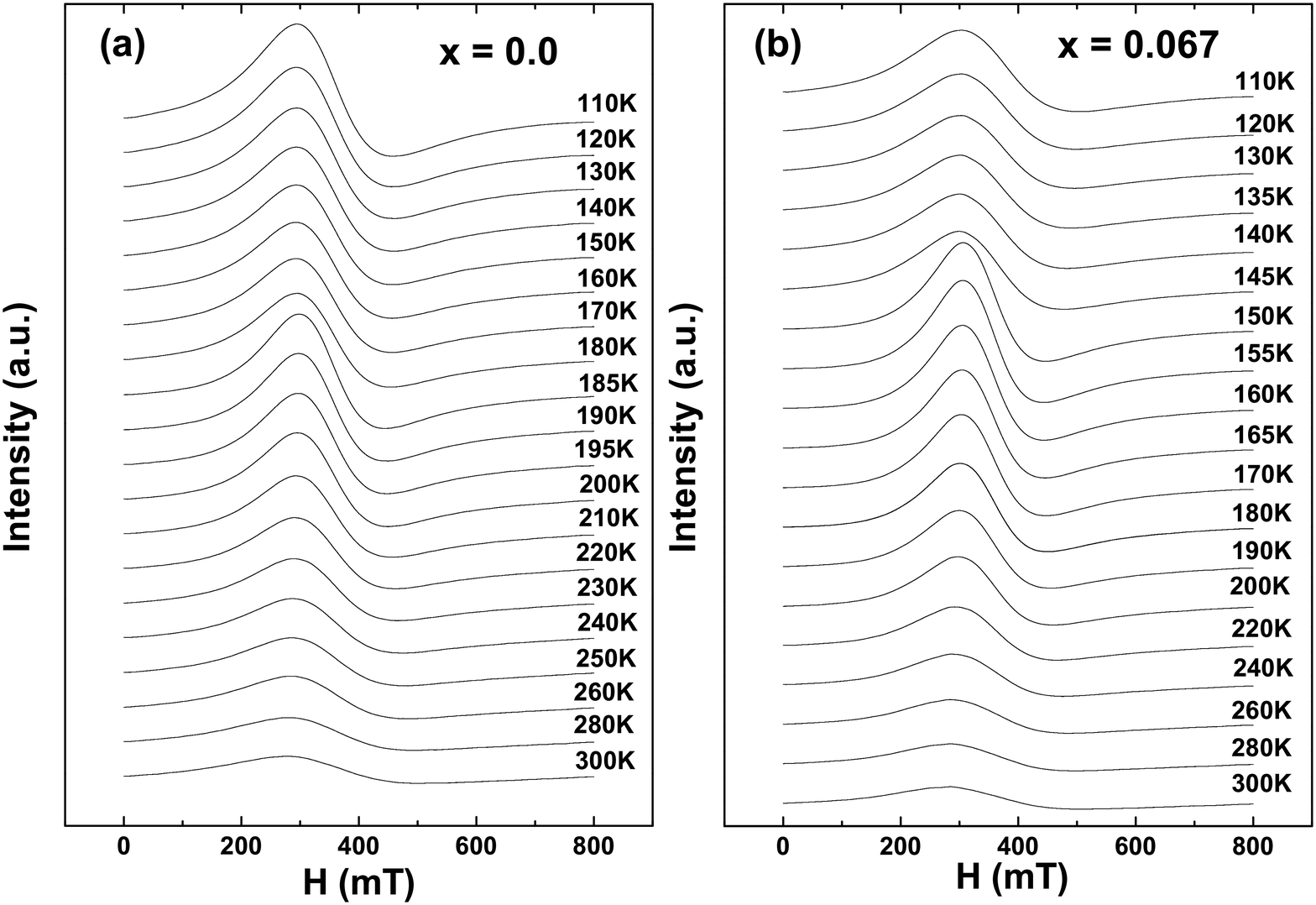}
\caption{Temperature dependence of ESR spectrum for (a): x=0.0 and
(b): x=0.067 . } \label{fig2}
\end{figure}

We analyzed the temperature dependence of ESR spectrum for x=0.0 and
x=0.067 samples at the temperature ranging from 110 K to 300 K. For
the two samples, the SDW transition temperatures are 188 K and 158 K
determined from the resistivity as shown in Fig.1, respectively.
Fig.3(a) illustrates the peak-to-peak linewidth ($\Delta$H$_{pp}$)
and the g factor as a function of temperature for the x=0.0 sample.
$\Delta$H$_{pp}$ is defined as the width between the highest point
and the lowest point in the temperature dependent ESR spectrum. The
linewidth decreases linearly with decreasing temperature above the
SDW transition temperature. This behavior is also observed in the
isostructural $EuCu_2Si_2$ single crystals\cite{Cao}.
$\Delta$H$_{pp}$ shows a sudden jump and begins to deviate from the
linear trend at $T_{SDW}$. At the temperature below SDW transition,
$\Delta$H$_{pp}$ becomes independent on temperature. The resonance
field (Hc) to calculate the effective g factor is defined as the
magnetic field corresponding to the midpoint between the highest and
lowest points in the ESR spectrum. The effective g factor is
calculated by the following formula: $g=\frac{h\nu}{\mu_BH_c}$. For
the parent compound, the g factor monotonously increases with
decreasing temperature, and a kink is observed at $T_{SDW}$. The ESR
intensity of the parent compound obtained by numerical integration
is illustrated in fig.3(b). It is found that the ESR intensity
follows Curie-Weiss formula at high temperature with a kink at
$T_{SDW}$. It is well known that ESR intensity is proportional to
spin susceptibility, and the behavior of spin susceptibility is very
much alike of DC susceptibility. This observation indicates that the
same magnetic species contributes to the ESR spectrum as the dc
susceptibility. It should be addressed that the drop of DC
susceptibility at $T_{SDW}$ cannot be observed because of the large
magnetic moment of $Eu^{2+}$,\cite{wu} while the anomaly at
$T_{SDW}$ in spin susceptibility is very obvious as shown in
Fig.3(b). Fig.3(c) and (d) show the temperature dependence of the
linewidth, the g factor and the ESR intensity of the x=0.067 sample.
The behavior of these three parameters is similar to that of the
parent compounds, and an anomaly is observed at $T_{SDW}$. The
$\Delta$H$_{pp}$ also shows the linear behavior above $T_{SDW}$. All
these results confirm the interaction between the local moments of
the $Eu^{2+}$ and the conduction electrons in the FeAs layers.
Comparing the x=0.067 sample with the parent compound, we find the
jump of the linewidth at $T_{SDW}$ is larger than the parent
compound. Moreover, the drop of ESR intensity at the $T_{SDW}$ for
the x=0.067 sample is much sharper than the parent compounds.

The linewidth and the g factors of four typical Co doped crystals
with x=0.20, 0.25, 0.285 and 0.40 are shown in Fig.4. As shown in
Fig.1, the SDW transition temperature was suppressed down to 58 K
for the x=0.20 crystal. The x=0.25 crystal with $T_c=22$ K is in the
underdoped region. The x=0.285 sample is at the optimal doping level
and its $T_c$=24.5 K, nearly the same as the highest $T_c$=25 K in
Co-doped $BaFe_2As_2$.\cite{wang2} The x=0.40 sample with $T_c$=20 K
is in the overdoped region. Strong temperature dependence of
linewidth and the g factors are observed in all the samples. The
linewidth of all the samples follows the linear temperature
dependence. This phenomenon is consistent with the behavior observed
in the x=0 and 0.067 samples above $T_{SDW}$ as shown in Fig.3(a)
and (c). The g factor of all the four samples increases monotonously
with decreasing temperature. No anomalies are found in these four
compounds.

\begin{figure}[t]
\centering
\includegraphics[width=0.5\textwidth]{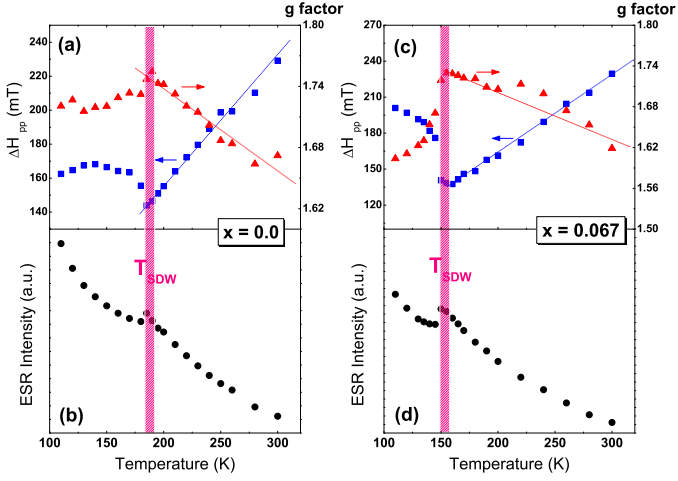}
\caption{(color online). Temperature dependence of g factor and
$\triangle$H$_{pp}$ for the crystal with (a): x = 0.0 and (c): x =
0.067. Temperature dependence of ESR intensity for the crystal with
(b): x = 0.0 and (d): x = 0.067. The blue line shows the fitting
with linear behavior.} \label{fig3}
\end{figure}

In order to systematically investigate the effect of Co doping in
$EuFe_{2-x}Co_xAs_2$, the ESR spectrum of the other samples with
different doping level was measured. It is found that the
temperature dependence of linewidth for all the samples shows the
same behavior. The linear temperature dependent linewidth is
observed in all the samples above $T_{SDW}$. Figure 5 illustrates
the slope of the linewidth, $T_{SDW}$ and $T_c$ as a function of Co
doping. It is striking that the slope of the linewidth shows a close
relation with the characteristic temperature $T_{SDW}$ and $T_c$. As
shown in Fig.5, the SDW ordering is gradually suppressed and
$T_{SDW}$ decreases with increasing Co doping. It is intriguing that
the slope of the linewidth shows the same behavior as that of
$T_{SDW}$, and the slope of the linewidth decreases with increasing
Co doping in the underdoped region. With further increasing the Co
doping, the superconductivity emerges and the slope of linewidth
begins to increase. In the superconducting region, the slope of
linewidth shows the same behavior as $T_c$. Both the slope of
linewidth and $T_c$ exhibit a dome-like behavior with increasing the
Co doping. The evolution of the slope of linewidth with Co-doping is
qualitatively consistent with that of $T_{SDW}$ and $T_c$.
\begin{figure}[t]
\centering
\includegraphics[width=0.5\textwidth]{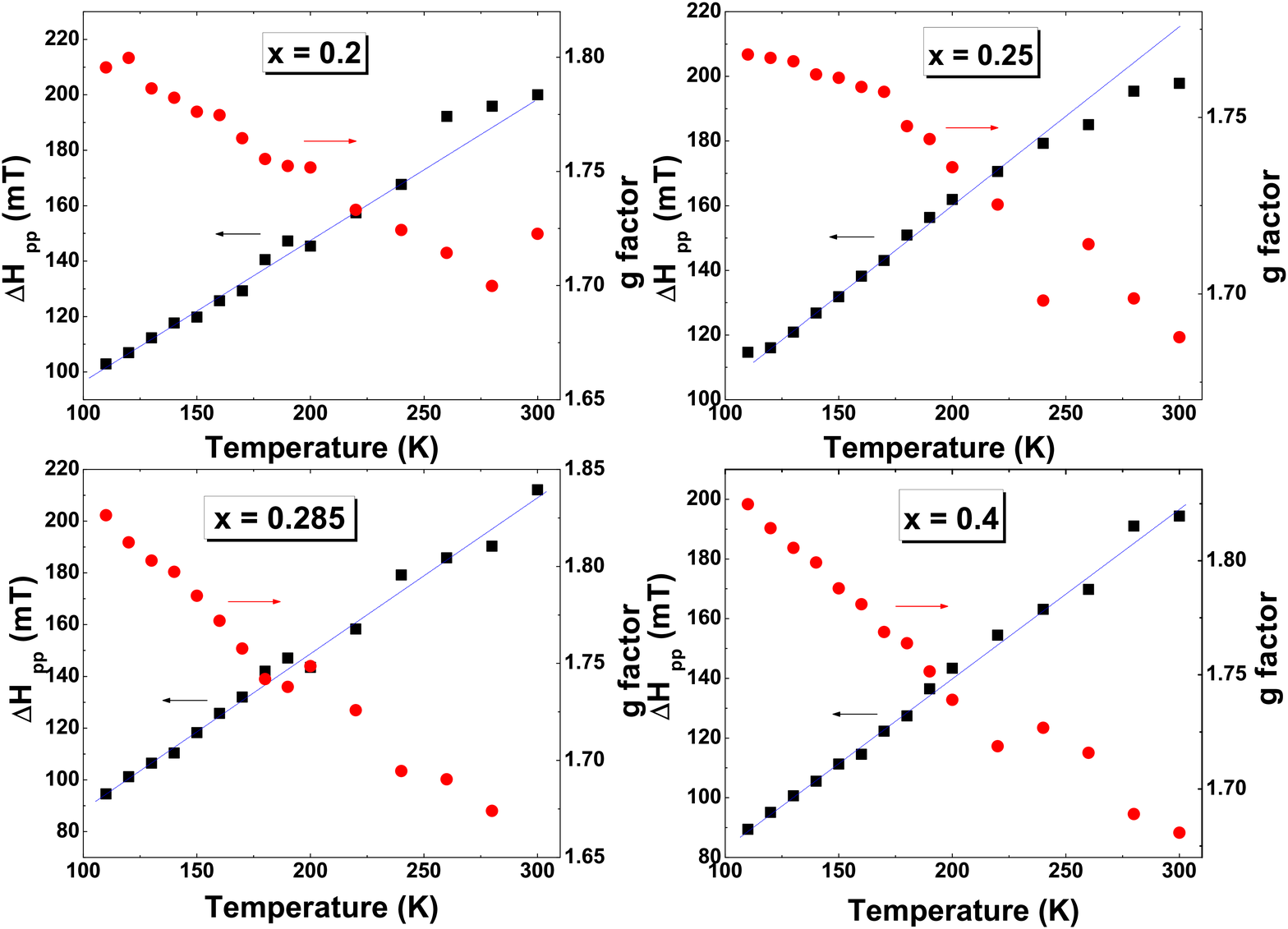}
\caption{(color online). Temperature dependence of g factor and
$\triangle$H$_{pp}$ for the crystal with (a): x=0.20; (b): x=0.25;
(c): x=0.285 and (d): x=0.4. The blue line shows a linear fitting. }
\label{fig4}
\end{figure}
Spin relaxation of magnetic moments in conventional metals mainly
depends on the relaxation times between local moments and itinerated
electrons, and spin-lattice relaxation time for itinerated electrons
and spin-lattice relaxation time for local
moments\cite{Taylor,Barnes,wutao}. The linewidth is a measurement of
the relaxation for the spin system. The exchange interaction between
local moments and conduction electrons results in a
temperature-linear contribution to the total ESR
linewidth\cite{shaltiel}. It follows the Korringa behavior
$\Delta$H=$\Delta$H$_K$+$\Delta$H$_{res}$. $\Delta$H$_{res}$ is the
residual linewidth and $\Delta$H$_K$ is the Korringa broadening
given by $\Delta$$H_K$ = $\frac{\pi K_BT}{g \mu _B}$
($J_{Ss}$$\rho$($\epsilon_F$))$^2$. Here $\mu_B$ is the Bohr
magneton, $K_B$ is the Boltzmann constant, $J_{Ss}$ is the exchange
interaction between the local moments (S) and the spin of the
conduction carriers (s), $\rho$($\epsilon_F$) is the density of
states at Fermi surface. The local moment should come mainly from
the Eu atom in $EuFe_{2-x}Co_{x}As_2 $ system. The linear linewidth
above $T_{SDW}$ for all the samples follows the Korringa law. When
SDW transition happens, the linear linewidth behavior is destroyed.
The 'free-ion' g factor for Eu is normally given as 1.993 in an
insulating host\cite{Barnes}. The g factor of Eu ions in these
materials is much smaller than 1.993 in the whole temperature range.
The shift of g factor is very sensitive to the temperature, and a
kink shows up at the temperature corresponding to SDW transition.
This behavior is different with the LaOFeAs system in which
ferromagnetic fluctuation from local moment of Fe ions was
observed\cite{wutao}. Such difference could be related to the large
local moments of $Eu^{2+}$. The ESR integral intensity, which is
proportional to the spin susceptibility, also shows anomalies at
$T_{SDW}$. In the parent compounds the linewidth becomes independent
on the temperature below $T_{SDW}$, so the linewidth slope is zero.
If it is assumed that the Korringa law is applicable below
$T_{SDW}$, then the $\rho$($\epsilon_F$) must be zero. This result
means that the density of state losses at the Fermi surface and a
gap opens at the temperature below $T_{SDW}$. Such behavior of
linewidth has been observed in the Kondo insulator
CeNiSn\cite{mair}. The slope of linewidth exhibits an interesting
behavior with the Co doping, and it shows a close relation with the
characteristic temperature $T_{SDW}$ and $T_c$. Because the $J_{Ss}$
between the $Eu^{2+}$ and the conduction electrons does not change
with Co doping, the difference of the slope must come from the
density of state at Fermi surface. Recently, Terashima et
al.\cite{Terashima} observed isotropic SC gaps which depend on the
nesting conditions of Fermi surface in the iron-based
superconductors. The superconductivity in this system is related to
the Fermi surface topology\cite{sekiba}. The Fermi surface nesting
maybe play an important part in the superconductivity, and the
nesting of Fermi surface has a strong effect on the slope of the
linewidth. There are two possible explanation for the SDW ordering.
One explanation is that the Fermi surface is perfectly nesting for
the parent compounds, resulting in SDW ordering. With Co doping, the
Fermi surface nesting is gradually destroyed, so that the slope
decreases with the Co doping in the SDW region. While in the SC
region, the Fermi surface nesting condition is changed, and the
superconductivity is closely related to the Fermi surface nesting
based on the study by Terashima et al.\cite{Terashima}. The behavior
of the slope associated with the $T_{SDW}$ and $T_c$ could be
related to the complicated nesting of Fermi surface.

\begin{figure}[t]
\centering
\includegraphics[width=0.5\textwidth]{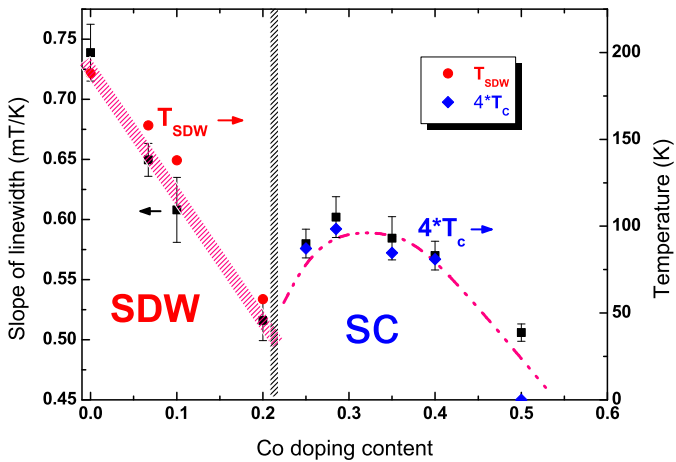}
\caption{(color online). The slope of peak-to-peak linewidth as a
function of the Co doping content. The black squares denote the
slope of linewidth, and the red circles present the SDW transition
temperature and the blue diamonds presents the 4*$T_C$.}
\label{fig5}
\end{figure}

In conclusion, we studied the temperature dependence of ESR spectrum
for single crystals $EuFe_{2-x}Co_{x}As_2$ (x=0 ,0.067 ,0.1 ,0.2
,0.25 ,0.285 ,0.35 ,0.4 and 0.5). Strong temperature dependence of
the g factor and $\Delta$H$_{pp}$ are observed in all the samples.
The linewidth, the g factor and the ESR intensity all show anomalies
at the SDW temperature. The linewidth for all the samples above
$T_{SDW}$ shows the Korringa behavior. These results indicate the
exchange interaction between the local moments of $Eu^{2+}$ and the
conduction electrons in FeAs layers. A gap opening at the SDW
transition is evidenced by the fact that the linewidth does not
change below the SDW temperature. The slope of the linewidth is
strongly dependent on the Co doping content, and is closely
associated with $T_{SDW}$ and $T_C$. Such behavior maybe related to
the complicated Fermi surface nesting.

This work is supported by the Nature Science Foundation of China
and by the Ministry of Science and Technology of China (973
project No: 2006CB601001) and by National Basic Research Program
of China (2006CB922005).


\begin{references}

\bibitem{yoichi}
Y. Kamihara, T. Watanabe, M. Hirano, and H. Hosono, \emph{J. Am.
Chem. Sco.} {\bf 130}, 3296(2008).
\bibitem{chenxh}
X. H. Chen and T. Wu, G. Wu, R. H. Liu, H. Chen and D. F. Fang,
Nature {\bf 453}, 761(2008).
\bibitem{chen}
G. F. Chen, Z. Li, D. Wu, G. Li, W. Z. Hu, J. Dong, P. Zheng, J.
L. Luo, and N. L. Wang, Phys. Rev. Lett. {\bf 100}, 247002(2008).
\bibitem{ren}
Z. A. Ren, G. C. Che, X. L. Dong, J. Yang, W. Lu, W. Yi, X. L.
Shen, Z. C. Li, L. L. Sun, F. Zhou and Z. X. Zhao, Europhys. Lett.
{\bf 83}, 17002(2008).
\bibitem{rotter}
M. Rotter, M. Tegel, D. Johrendt, Phys. Rev. Lett. {\bf 101},
107006(2008).
\bibitem{zhi ren}
Zhi Ren, Zengwei Zhu, Shuai Jiang, Xiangfan Xu, Qian Tao, Cao Wang,
Chunmu Feng, Guanghan Cao and Zhu¡¯an Xu, Phys. Rev. B {\bf 78},
052501 (2008).
\bibitem{wu}
T. Wu, G. Wu, H. Chen, Y. L. Xie, R. H. Liu, X. F. Wang and X. H.
Chen, arXiv:0808.2247. J. Mag. Mag. Mat. (in press).
\bibitem{shaltiel}
D. Shaltiel, C. Noble, J. Pilbrow, D. Hutton, E. Walker, Phys. Rev.
B {\bf 53}, 12430 (1996).
\bibitem{kataev}
V. Kataev, Yu. Greznev, G. Teitel'baum, M. Breuer and N. Knauf,
Phys. Rev. B {\bf 48}, 13042 (1993).
\bibitem{wang}
X. F. Wang, T. Wu, G. Wu, H. Chen, Y. L. Xie, J. J. Ying, Y. J. Yan,
R. H. Liu, X. H. Chen,  Phys. Rev. Lett. {\bf 102}, 117005(2009).
\bibitem {zheng}
Q. J. Zheng, Y. He, T. Wu, G. Wu, H. Chen, J. J. Ying, R. H.~Liu, X.
F. Wang, Y. L. Xie, Y. J. Yan, Q. J. Li and X. H. Chen, unpublished.
\bibitem{Cao}
C. D. Cao, R. Klingeler, N. Leps, H. Vinzelberg, V. Kataev, F.
Muranyi, N. Tristan, A. Teresiak, Shengqiang Zhou, W. L\"{o}ser, G.
Behr and B. B\"{u}chner, Phys. Rev. B {\bf 78}, 064409 (2008).
\bibitem{wang2}
X. F. Wang, T. Wu, G. Wu, R. H. Liu, H. Chen, Y. L. Xie, X. H. Chen,
New J. Phys. {\bf 11}, 045003 (2009).

\bibitem{Taylor}
R. H. Taylor, Advances in Physics {\bf 24}, 681-791 (1975).
\bibitem{Barnes}
S. E. Barnes, Advances in Physics {\bf 30}, 801-938 (1981).
\bibitem{wutao}
T. Wu, J. J. Ying, G. Wu,  R. H. Liu, Y. He, H. Chen, X. F. Wang, Y.
L. Xie, Y. J. Yan, and X. H. Chen, Phys. Rev. B {\bf 79}, 115121
(2009).
\bibitem{mair}
S. Mair, H.-A. Krug von Nidda, M. Lohmann and A. Loidl, Phys. Rev. B
{\bf 60}, 16409 (1999).
\bibitem{Terashima}
 K. Terashima, Y. Sekiba, J. H. Bowen, K. Nakayama, T. Kawahara, T. Sato, P. Richard, Y.-M. Xu, L. J. Li, G. H. Cao, Z.-A. Xu, H. Ding, T.
 Takahashi,
Proceedings of the National Academy of Sciences of the USA (PNAS)
{\bf 106}, 7330-7333 (2009)
\bibitem{sekiba}
Y. Sekiba, T. Sato, K. Nakayama, K. Terashima, P. Richard, J. H.
Bowen, H. Ding, Y.-M. Xu, L. J. Li, G. H. Cao, Z.-A. Xu, T.
Takahashi, New J. Phys {\bf 11}, 025020 (2009).



\newpage

\noindent

\end{references}
\end{document}